\documentclass[12pt]{iopart}
\usepackage{iopams}  
\usepackage[utf8]{inputenc}
\usepackage[dvips]{graphicx}
\usepackage{amsthm}
\expandafter\let\csname equation*\endcsname\relax
\expandafter\let\csname endequation*\endcsname\relax
\usepackage{amsmath}
\usepackage{amssymb}
\usepackage{mathrsfs} 
\usepackage[sort&compress,numbers]{natbib}
\usepackage{enumerate}
\usepackage[dvips]{graphicx}
\usepackage[usenames,dvipsnames]{xcolor}
\usepackage[hyperindex,breaklinks]{hyperref}
\usepackage{appendix}
\usepackage{color}
\newcommand{\ud}{\mathrm{d}}
\newcommand{\ue}{\mathrm{e}}

\begin{document}

\title{The model of the local Universe in the framework of the second-order perturbation theory}
\author{Szymon Sikora$^1$, Jan J. Ostrowski$^2$}
	\address{$^1$ Astronomical Observatory, Jagiellonian University, Orla 171, 30-244 Kraków, Poland}
 \ead{szymon.sikora@uj.edu.pl}
	%\and
%	\author{Jan J. Ostrowski}  
\vspace{5pt}
	\address{$^2$ Department of Fundamental Research, National Centre for Nuclear Research, Pasteura 7, 02–093 Warsaw, Poland}
\ead{Jan.Jakub.Ostrowski@ncbj.gov.pl}
\vspace{10pt}
\begin{indented}
\item[]June 2024
\end{indented}

	\begin{abstract}
		Recently, we constructed the specific solution to the second-order cosmological perturbation theory, around any Friedmann–Lema\^itre–Robertson–Walker (FLRW) background filled with dust matter and a positive cosmological constant. In this paper, we use the {\it Cosmicflows-4} (CF4) sample of galaxies from the Extragalactic Distance Database to constrain this metric tensor. We obtain an approximation to the local matter distribution and geometry. We numerically solve for null geodesics for randomly distributed mock sources and compare this model with the Lema\^itre-Hubble constant inferred from the observations under the assumption of perfect isotropy and homogeneity. We conclude on effects of realistic inhomogeneities on the luminosity distance in the context of the Hubble tension and  discuss limitations of our approach.
	\end{abstract}
%	\vspace{2pc}
%\keywords{XXXXXX, YYYYYYYY, ZZZZZZZZZ}
%\submitto{\CQG}

%\maketitle

	\section{Introduction}
	%general intro
	
 Persistent discrepancies between the  Cosmic Microwave Background (CMB)-inferred and locally measured values of the Lema\^itre-Hubble constant have elicited plenty of proposals of solutions, none of which however,  seems to be fully convincing so far (see e.g. \cite{Hubbletrouble} for a review).  Majority of the suggested solutions to the Hubble tension involves extensions of the $\Lambda$ CDM model but retaining the {\it Cosmological Principle} i.e. the assumption that the universe is isotropic and homogeneous on large enough scales. Another significant tension arising from the difference in the amplitude of the CMB dipole and the dipole in the local distribution of sources \cite{2021Secrest} puts the assumption of the isotropy of the universe to question.
	%inhomogeneous cosmology
	Within the inhomogeneous cosmology approach the aim is to examine the effects of the energy density and curvature inhomogeneities on the light propagation. This can be done either by studying exact solutions or using perturbation theory. In both cases, the luminosity distance -- redshift relation reveals that the cosmological parameters are `dressed' when interpreted strictly within the FLRW model. Specifically, additional, physical degrees of freedom generically present in the inhomogeneous models affect the what-would-be Hubble constant or deceleration parameter in the FLRW model. Significantly restricted number of free parameters in the perfectly homogeneous and isotropic universe has to be appended by kinematic and dynamical quantities like: shear, local expansion, acceleration and certain projections of Riemann curvature to name the few (\cite{2003BuchertCarfora}, \cite{Asta2021}). To make the problem more tractable one needs to make some physically sound assumptions on the geometry and matter content which can be obtained either by symmetries or quasi-symmetries of the space-time in the case of exact solutions or the perturbative character of the model. In this work we take the latter approach and apply the second-order cosmological perturbation theory to firstly: model the local distribution of matter smoothed at certain scale and secondly: calculate the effect of the perturbations on the relevant observables. 

	In a series of papers, we considered specific solutions to the cosmological perturbation theory. In \cite{2017PhRvD..95f3517S} we constructed an infinite cubic lattice of inhomogeneities by performing linear perturbations around the Einstein-de Sitter background. Since the metric tensor was explicitly given, we were able to apply the Green-Wald \cite{2011GreenWald} and the Buchert averaging \cite{2000Buchert} schemes and analyze how the light propagates through the spacetime before and after application of the respective averaging procedure. Then, the model was generalized beyond the linear order in \cite{2019PhRvD..99h3521S}. In both cases the amplitude of inhomogeneities was a decreasing function of time. The growing mode was studied in \cite{2021EPJC...81..208S}. Finally, in \cite{2023CQGra..40b5002S} we presented the most general case, for which perturbations up to the second-order are performed around any FLRW background spacetime filled with dust, and for a possibly non-vanishing cosmological constant. The aim of the present work is to constraint this model with the local matter distribution and confront it with the observations. 
	
	In the model presented in \cite{2023CQGra..40b5002S}, there is a freedom of choosing one of the metric functions, called $C_{10}$. This function is directly related to the spatial distribution of matter. In Section \ref{Sec:Setup}, we show a possible way of constraining the metric function $C_{10}$ based on the observed distribution of nearby galaxies. When the function $C_{10}$ is given, the metric tensor depends only on the background parameters and the parameter controlling the amplitude of the inhomogeneities. In Section \ref{Sec:Observables}, we focus on the Lema\^itre-Hubble constant associated with the presented inhomogeneous metric and discuss the shortcomings of our approach. We summarize our findings in the last Section \ref{Sec:Conclusions}.     
	
	\section{The setup}\label{Sec:Setup}
	\subsection{The metric tensor}
	Let us begin by recalling the most important formulas from the previous work \cite{2023CQGra..40b5002S}. The perturbed FLRW line element can be written in the Cartesian-like coordinates $x^\mu=(t,x,y,z)$ as:
	\begin{equation}\label{eqn:Metric1}
		\ud s^2=-\ud t^2+a(t)^2\,\sum\limits_{l=0}^2\sum\limits_{m=0}^{l}c_{i\:\!j}^{(l-m,m)}(x^\mu)\,\lambda^{l-m}\,k^m\,\ud x^i\,\ud x^j\,.
	\end{equation}
	Here, $a(t)$ is the scale factor of the background, $c_{i\:\!j}^{(l,m)}$ are the coordinate functions labeled with the indices $(l,m)$, $\lambda$ is the parameter proportional to the amplitude of the inhomogeneities, and $k$ is the curvature parameter. In the limit $\lambda\rightarrow 0$, the line element reduces to:
	\begin{equation}\label{eqn:background_g}
		\ud s^2=-\ud t^2+a(t)^2\,\left(1-\frac{1}{2}\,k\,\widetilde{r}\,{}^2+\frac{3}{16}\,k^2\,\widetilde{r}\,{}^4 \right)\,\delta_{i\:\!j}\,\ud x^i\,\ud x^j\,,
	\end{equation} 
	where $\delta_{i\:\!j}$ is the Kronecker delta symbol, and $\widetilde{r}\,{}^2=x^2+y^2+z^2$. The metric (\ref{eqn:background_g}) is the Taylor expansion up to the second order in $k$ of the FLRW metric expressed in the Cartesian-like coordinates:
	\begin{equation}\label{eqn:FLRW_cartesian}
		\ud s^2=-\ud t^2+a(t)^2\,\frac{\delta_{i\:\!j}}{\left(1+\frac{1}{4}k\,(x^2+y^2+z^2) \right)^2}\,\ud x^i\,\ud x^j.
	\end{equation}
	The FLRW metric is usually written in the spherical coordinates $(t,r,\theta,\phi)$:
	\begin{equation}
		\ud s^2=-\ud t^2+a(t)^2\left(\frac{\ud r^2}{1-k\,r^2}+r^2\,\ud\Omega^2 \right)
	\end{equation}
	Coordinate transformation to the Cartesian-like coordinates has the following form: $x=\widetilde{r}\,\sin\theta\,\cos\phi$, $y=\widetilde{r}\,\sin\theta\,\sin\phi$, $z=\widetilde{r}\,\cos\theta$. For a spatially flat spacetime $k=0$ radial variables are the same $\widetilde{r}=r$. When $k\neq 0$ the radial variable $\widetilde{r}=2\,(1-\sqrt{1-k\,r^2})/(k\,r)$ is compactified $\widetilde{r}\in[0,2/\sqrt{|k|}]$. In this article we use natural units $c=1$ and $G=1$, and a megaparsec as a unit of length. For an exemplary values of the background parameters $1-\Omega_m-\Omega_\Lambda=0.1$ and the Lema\^itre-Hubble constant $H_0=67\,  \mathrm{km/s/Mpc}$ the curvature parameter's value is around $k\approx 10^{-8}\,\mathrm{Mpc}^{-2}$, and $2/\sqrt{k}\approx 3\times 10^4\,\mathrm{Mpc}$. 
 
 The presented model can be used to describe a local neighborhood of the observer, where $\widetilde{r}$ is small enough so that the metric (\ref{eqn:background_g}) approximates well the FLRW background (\ref{eqn:FLRW_cartesian}). We will investigate a matter periodically distributed within a cubic lattice. For an elementary cell of the size $L=60\,\mathrm{Mpc}$ this condition is satisfied and a difference between the coordinate values $r$ and $\widetilde{r}$ is negligible. Moreover, for the object with the luminosity distance $d_L=60\,\mathrm{Mpc}$, the difference between $\widetilde{r}$ and $d_L$ is of the order of $10^{-3}\,\mathrm{Mpc}$. For that reason, the coordinates $(x,y,z)$ can be understood as the supergalactic coordinates $(SGX,SGY,SGZ)$.
	
	We propose the following form of the first-order correction to the FLRW metric as a diagonal matrix:
	\begin{equation}
		%	c_{i\:\!j}^{(1,0)}=\mathscr{A}_{10}(t)\frac{\partial^2}{\partial x^i \partial x^j}(\left C_{10}(x)+C_{10}(y)+C_{10}(z)\right)+\alpha\,\left(C_{10}(x)+C_{10}(y)+C_{10}(z)\right)\,\delta_{i\:\!j}\,.
		\left(c_{i\:\!j}^{(1,0)}\right)=\mathrm{diag}\left(\mathscr{A}_{10}(t)C_{10}''(x)+B_{10}\,,\,\mathscr{A}_{10}(t)C_{10}''(y)+B_{10}\,,\,\mathscr{A}_{10}(t)C_{10}''(z)+B_{10} \right)\,,
	\end{equation}
	where $B_{10}\equiv \alpha(C_{10}(x)+C_{10}(y)+C_{10}(z))$, the prime denotes the derivative $C_{10}'(x)\equiv {\ud C_{10}(x)}/{\ud x}$, and $\alpha$ is a constant. The function of time $\mathscr{A}_{10}(t)$ should satisfy the following differential equation:
	\begin{equation}\label{eqn:A1}
		%a^2\,\ddot{\mathscr{A}_{10}}+3\,a\,\dot{a}\,\dot{\mathscr{A}_{10}}=\alpha\,,
		a^2(t)\,\ddot{\mathscr{A}}_{10}(t)+3\,a(t)\,\dot{a}(t)\,\dot{\mathscr{A}}_{10}(t)=\alpha\,.
	\end{equation} 
	Since the scale factor of the background is known, the function $\mathscr{A}_{10}(t)$ can be found with the help of the variation of constants method and in the integral form the solution reads:
	\begin{eqnarray}\label{eqn:A10}
		\mathscr{A}_{10}(t)=\int\limits_{t_i}^{t}f_1(t')\,\ud t'+\mathcal{C}_1, \quad P_1(t)=3\int\limits_{t_i}^{t}\frac{\dot{a}(t')}{a(t')}\,\ud t'\\
		\nonumber \mathrm{and}\:\:f_1(t)=\left(\mathcal{C}_2+\alpha\int\limits_{t_i}^{t}\frac{\ue^{P_1(t')}}{a^2(t')}\,\ud t' \right)\,\ue^{-P_1(t)}\,.
	\end{eqnarray}
	Ignoring the integration constants $\mathcal{C}_1=\mathcal{C}_2=0$ and taking the initial time close to the Big Bang $t_i\approx 10^{-3}$, those integrals can  be easily  evaluated numerically.
	For the metric functions specified above, the Einstein equations up to the first order in $\lambda$ are satisfied by the energy momentum tensor of the dust $T_{\mu\:\!\nu}=\rho\,U_\mu\,U_\nu$. The dust is comoving in the coordinates $(t,x,y,z)$, and its four-velocity $U^\mu=(1,0,0,0)$ is unperturbed. The energy density can be written as $\rho=\rho_0+\lambda \rho^{(1,0)}$, where $\rho_0=\rho_0(t)$ is the density of the FLRW background, and the first-order correction to the energy density is:
	\begin{equation}\label{eqn:density1}
		\rho^{(1,0)}(t,x,y,z)=\frac{a(t)\,\dot{a}(t)\,\dot{\mathscr{A}}_{10}(t)-\alpha}{8\pi\,a^2(t)}\,\left(C_{10}''(x)+C_{10}''(y)+C_{10}''(z) \right)\,.
	\end{equation}
	The function $\mathscr{A}_{10}(t)$ is proportional to $\alpha$, so a precise value of this constant plays no role. One can multiply $\alpha$ by an arbitrary factor and then redefine the parameter $\lambda$ by an inverse of this factor keeping the metric unchanged. For a practical reasons we use the freedom of choosing $\alpha$ to normalize the function $\mathscr{A}_{10}(t_0)=1$ at the universe age $t_0$. 
	
	For the reader convenience, we moved to the \ref{sec:2d_order_metric} the explicit form of the second-order metric tensor, for which the Einstein equations up to the second order are approximately satisfied by the energy-momentum tensor of the dust, with the following energy density: 
	$\rho=\rho_0+\lambda \rho^{(1,0)}+\lambda\,k\,\rho^{(1,1)}+\lambda^2\,\rho^{(2,0)}$.
	At this point, it is sufficient to notice that the metric tensor up to the second-order is completely determined by the background parameters $(\Omega_m,\Omega_\Lambda,H_0)$, which provide us the scale factor $a(t)$ and the value of the curvature parameter $k$, the value of the amplitude $\lambda$, and the arbitrary metric functions $C_{10}(x)$, $C_{10}(y)$ and $C_{10}(z)$. In the work \cite{2023CQGra..40b5002S}, these three functions have the same form. In general, the functions $C_{10}$ of different spatial variables could be understood as a three distinct functions.
	In the next subsection, we show how the functions $C_{10}$ can be determined based on the observed distribution of nearby galaxies.
	
	\subsection{The density distribution profile}\label{Sec:dens_distr}
	It is a reasonable assumption that the probability of finding a galaxy in the volume $\ud x\,\ud y\,\ud z$, at the universe's age $t_0$, is proportional to the density fluctuations at that time. In the first order of the perturbation theory, this condition reads:
	\begin{equation}
		f(x,y,z)\,\ud x\,\ud y\,\ud z \propto \rho^{(1,0)}(t_0,x,y,z)\,\ud x\,\ud y\,\ud z\,.
	\end{equation}
	Then, the iso-density surfaces of $\rho^{(1,0)}$ are the same as the iso-surfaces of the probability distribution $f$, but for a different constants. Let us assume the following form of the probability distribution $f$, which is consistent with the structure of the first-order density (\ref{eqn:density1}): 
	\begin{eqnarray}
		\label{eqn:probability_f}f(x,y,z)=\frac{1}{\mathcal{N}}\sum_{i=1}^{S}A_i\Big(d_i(x,X_i,a_i)+d_i(y,Y_i,a_i)+d_i(z,Z_i,a_i) \Big)\,,\\
		\label{eqn:di}d_i(x,X_i,a_i)=\exp\left(a_i\,\cos\left( \frac{\pi}{L}(x-X_i)\right)\right)\,,\\
		\mathcal{N}=3\,(2L)^3\,\sum_{i=1}^{S}A_i\,I_0(a_i)\,.
	\end{eqnarray}
	The functions $d_i(y,Y_i,a_i)$ and $d_i(z,Z_i,a_i)$ have the  form analogical to (\ref{eqn:di}), $I_n(x)$ is the modified Bessel function of the first kind, and $\{a_i,A_i,X_i,Y_i,Z_i\}$ are the parameters. The iso-surfaces of $f$ form a periodic, cubic lattice, with a period $2L$. Within the elementary cell $x,y,z\in [-L,L]$ the function $f$ has properties of a probability distribution: it is non-negative and its integral over the elementary cell is equal to one. The considered probability distribution $f$ accounts for the $S$ substructures within the elementary cell. Each substructure is characterized by the its amplitude $A_i$, position of its density peak $(x,y,z)=(X_i,Y_i,Z_i)$ and the parameter $a_i$ describing the width of its density profile.
	
\begin{figure}[h]
		\centering
		\includegraphics[width=0.65\textwidth]{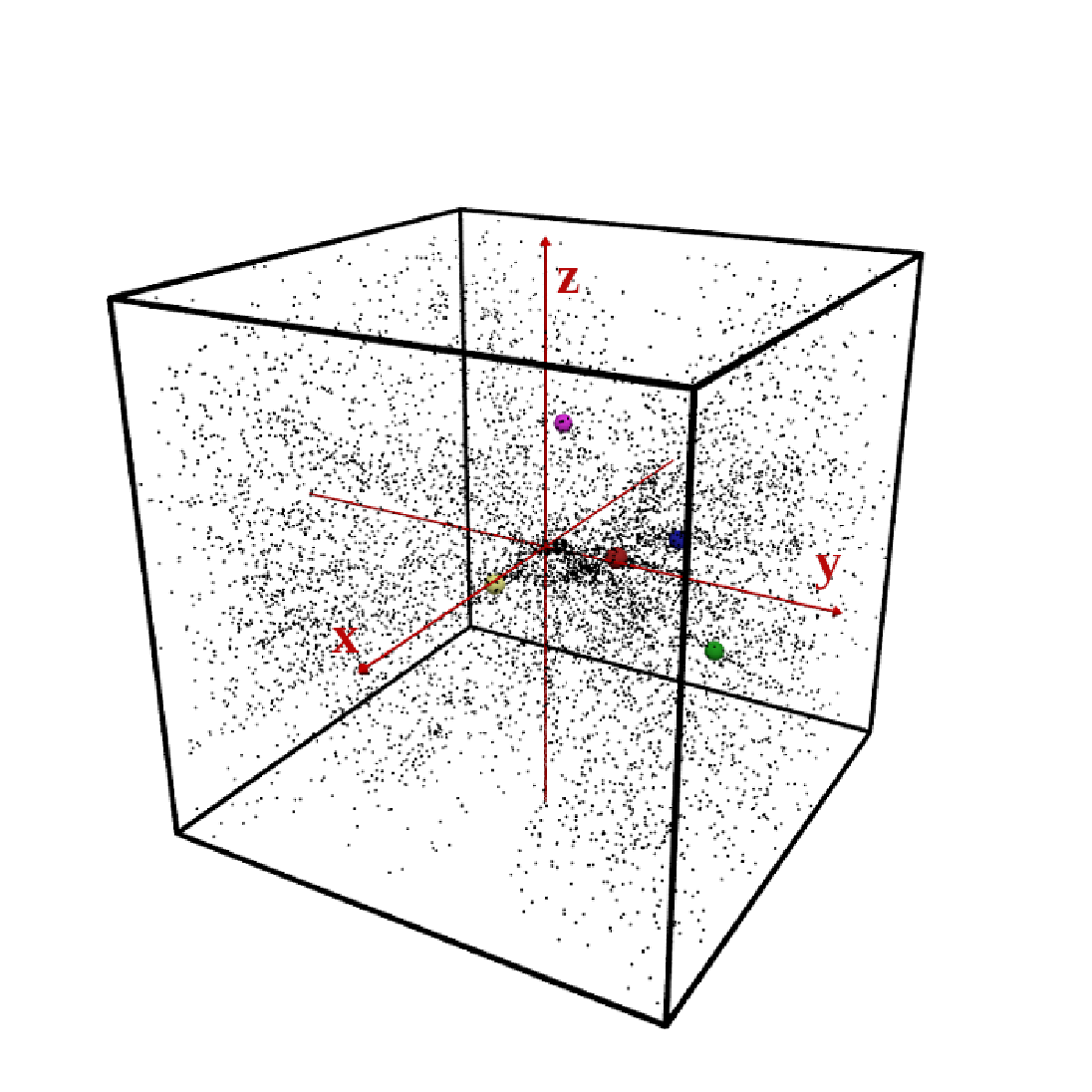}
		\caption{\label{fig:galaxies} \scriptsize{Distribution of the galaxies within the elementary cell $x,y,z\in[-L,L]$, with $L=60\,\mathrm{Mpc}$. \emph{Black points} - galaxies from the \emph{All CF4 Individual Galaxies}. Color points - positions of the centers $(X_i,Y_i,Z_i)$ of the galaxy clusters under consideration: \emph{red} - Virgo Cluster, \emph{green} - Hydra Cluster, \emph{yellow} - Formax Cluster, \emph{blue} - Centaurus Cluster, \emph{magenta} - Pavo Cluster.}}
	\end{figure}

	To derive the probability distribution function $f$ describing the local Universe, we use the sample of individual galaxies \emph{CF4} from \emph{The Extragalactic Distance Database} \cite{2023ApJ...944...94T,2009AJ....138..323T}. The richest galaxy clusters located at a distance from Earth smaller than $65\,\mathrm{Mpc}$ are: Virgo Cluster, Hydra Cluster, Fornax Cluster and Centaurus Cluster. In the recent study of the velocity field of galaxies \cite{2014Natur.513...71T} it was suggested that the our local supercluster of galaxies \emph{Laniakea} contains also the Pavo-Indus region. For that reason, in summation (\ref{eqn:probability_f}) we restrict ourselves to $S=5$ substructures: Virgo Cluster, Hydra Cluster, Fornax Cluster, Centaurus Cluster and additionally Pavo Cluster (S805). From the \emph{CF4 All Groups} data we derived their centroids $(X_i,Y_i,Z_i)$. They are shown in the Figure \ref{fig:galaxies}.
	
\begin{figure}[h]
		\centering
		\includegraphics[width=0.65\textwidth]{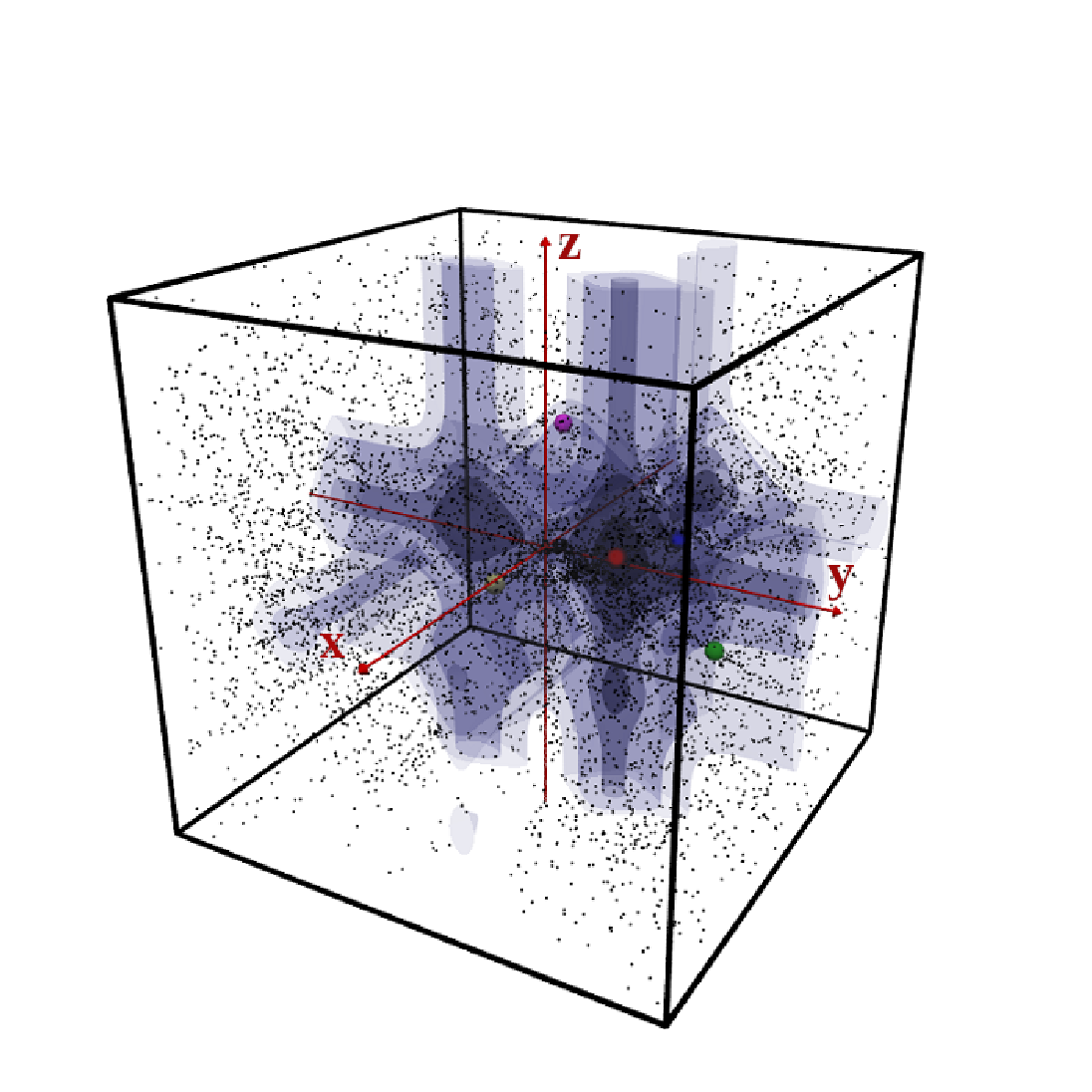}
		\caption{\label{fig:isosurfaces} \scriptsize{The isosurfaces of the probability distribution function $f$ for the parameters listed in the Table \ref{Tab:parameters}. }}
	\end{figure}

	We fix the amplitude $A_i$ of each substructure as the number of its member galaxies with optical surface brightness fluctuations (SBF) measurements  in the \emph{CF4 All Groups} database relative to the Virgo Cluster. We find the remaining parameters $a_i$ by minimizing the log-likelihood function for the probability distribution $f(x,y,z)$ evaluated at position of each galaxy from \emph{CF4 All Individual Galaxies} database within the elementary cell $x,y,z\in[-L,L]$, where $L=60\,\mathrm{Mpc}$. The minimization procedure provides the following values $a_1=9.75$, $a_2=9.48$, $a_3=9.67$, $a_4=0$, and $a_5=11.28$. For the value $a_4=0$ the Centaurus Cluster is not included in the density distribution. However, for the values $a_4\in[0,8]$ and the other parameters $a_i$ fixed, the log-likelihood function is almost flat. We may introduce the Centaurus Cluster with $a_4=8$ practically without changing the value of the log-likelihood function. All the parameters of the probability distribution $f$ are given in the Table \ref{Tab:parameters}. The iso-surfaces of the resulting probability distribution function $f$ are shown in the Figure \ref{fig:isosurfaces}.
	
	\begin{table}[h!]
		\centering	
		\begin{tabular}{| c c c c c c c|}
			\hline
			Name & N & $A_i$ & $a_i$ & $X_i$ & $Y_i$ & $Z_i$ \\
			\hline
			Virgo Cluster & 164 & 1.0 & 9.75 & -3.60 & 15.75 & -0.65 \\
			Hydra Cluster & 67 & 0.41 & 9.48 & -32.62 & 27.95 & -33.13 \\
			Fornax Cluster & 51 & 0.31 & 9.67 & -1.86 & -14.49 & -13.09 \\
			Centaurus Cluster & 49 & 0.30 & 8.0 & -36.11 & 15.83 & -8.08 \\
			Pavo Cluster & 18 & 0.11 & 11.28 & -49.79 & -23.49 & 10.83 \\
			\hline
		\end{tabular}
		\caption{\scriptsize{Parameters of the probability distribution function $f$. $N$ - number of galaxies with optical SBF measures in the \emph{CF4 All Groups} database. $A_i$ - the amplitude proportional to $N$. $a_i$ - the parameter controlling the width of the substructure peak. $(X_i,Y_i,Z_i)$ - cluster centroid position in the supergalactic coordinates, in the units of $\mathrm{Mpc}$.}}
		\label{Tab:parameters}
	\end{table}
	
	Now, we focus on the reconstruction of the metric functions  $C_{10}$ for a given probability distribution $f$. In (\ref{eqn:density1}) the term $(a(t)\,\dot{a}(t)\,\dot{\mathscr{A}}_{10}(t)-\alpha)/(8\pi\,a^2(t))$ is negative. The density $\rho^{(1,0)}$ and the probability distribution $f$ should be proportional to each other, and the constant of proportionality should be positive, so we might demand that:
	\begin{equation}
		C''_{10}(x)=-\sum\limits_{i=1}^{S} A_i\,d_i(x,X_i,a_i)\,.
	\end{equation}
	To obtain the metric function $C_{10}$ 
	one has to calculate the integral of $d_i(x,X_i,a_i)$ twice. Unfortunately, the indefinite integral of the function $\exp(\cos x)$ is not given by any elementary function. However, one can write $d_i(x,X_i,a_i)$ as a Fourier series on a domain $x\in[-L,L]$ and find a series expansion of the integral. Fourier series representation of $d_i$ reads:
	\begin{equation}
		d_i(x,X_i,a_i)=I_0(a_i)+2\sum\limits_{n=1}^{\infty}I_n(a_i)\,\cos\left(\frac{\pi}{L}n\,(x-X_i)\right)\,.
	\end{equation}
	If we neglect the constant $I_0(a_i)$, then the integral of the first-order density $\rho^{(1,0)}$ over the elementary cell is equal to zero, so the averaged density of the model tends to the density of the FLRW background, while the metric function $C_{10}$ is given by:
	\begin{equation}
		C_{10}(x)=2\,\sum_{i=1}^{S}\sum\limits_{n=1}^{\infty}I_n(a_i)\left(\frac{L}{n\,\pi}\right)^2\cos\left(n \frac{\pi}{L}(x-X_i)\right)\,.
	\end{equation}  
	The form of the solution for $C_{10}(y)$ and $C_{10}(z)$ is analogical. Because the value of the modified Bessel function of the first kind $I_n(a_i)$ decrease with growing $n$, only few first terms in the summation over $n$ can be taken into account to achieve desired accuracy. For the metric functions $C_{10}$ defined this way, the iso-surfaces of the probability density (\ref{eqn:probability_f}) and the iso-surfaces of the first-order density (\ref{eqn:density1}) are the same.
	
	As a result, we constructed the metric tensor of a perturbative model up to the second order, which is consistent with the observed distribution of galaxies at distances up to $L=60\,\mathrm{Mpc}$. We make here an assumption that this distribution is representative for an entire universe, and for distances larger than $L$ this density distribution is copied, forming an infinitive cubic lattice with a period $2L$. In the Figure \ref{fig:isosurfaces}, one can observe some artificial filaments in the density distribution, which are artifacts of the assumed symmetry of the metric. However, since the largest of such filaments coincides with the supergalactic plane and maxima of the density distribution corresponds to largest galaxy clusters centroids, we perceive the resulting density distribution as sufficiently realistic. 
	
	\section{The Hubble tension}\label{Sec:Observables}
	In this section, we investigate whether the inhomogeneous cosmological model presented in the previous section can fully or at least partially explain the difference between the CMB-inferred Lema\^itre-Hubble constant estimate and other estimates related to the low-redshift data. 
	
	Our approach is perturbative, with the density perturbations growing with time. At the epoch of recombination, the amplitude of the inhomogeneities in our model is negligible, and the perturbed metric tends to the background metric. To account for the CMB-based Lema\^itre-Hubble constant estimate from the Planck satellite \mbox{$H_0=67.4\,\mathrm{km/s/Mpc}$} \cite{2020A&A...641A...6P}, we fix this $H_0$ value for the background spacetime. We use it, along with the background parameters $\Omega_m$ and $\Omega_\Lambda$, to derive the curvature parameter $k=H_0^2\,(\Omega_m+\Omega_\Lambda-1)$, the scale factor $a(t)$, the universe age $t_0$, and the critical density $\rho_{cr}=(3H_0^2)/(8\pi)$. Then, we make use of the method presented in the previous section to get all the other metric functions, which carry an important contribution to the metric tensor in the late times. With the inhomogeneous model constructed this way, we show below a local Lema\^itre-Hubble constant estimates. 
	
	\subsection{The distance at the hypersurface of a constant time}
	In the background FLRW model, the Lema\^itre-Hubble constant is defined as the Hubble parameter $H(t)=\dot{a}(t)/a(t)$ evaluated at the universe's current age $H_0=H(t_0)$. For the object with spatial coordinates $(\widetilde{r},\theta,\phi)$, the physical distance to the origin at the hypersurface of a constant time $t=t_0$ is:
	\begin{equation}\label{eqn:dist}
		d=\int_{0}^{\widetilde{r}}\sqrt{g_{i\:\!k}\frac{\ud x^i}{\ud l}\,\frac{\ud x^k}{\ud l}}\,\ud l\,,
	\end{equation}
	where the possible parametrization of the line joining the object position with the observer at the origin in Cartesian-like coordinates is: 
	\begin{equation}
	x^\mu(l)=(t_0,l\,\cos(\theta)\cos(\phi),l\,\cos(\theta)\sin(\phi),l\,\sin(\theta))\,.
	\end{equation}
	In the FLRW background, the distance $d$ is proportional to the scale factor $a(t_0)$, so the Lema\^itre-Hubble constant value equals $\dot{d}/d$. In this subsection, we adopt the formula $H_0=\dot{d}/d$ as a definition of the Lema\^itre-Hubble constant, and apply it to the inhomogeneous model (\ref{eqn:Metric1}). The time derivative of the distance $d$ is:
	\begin{equation}\label{eqn:dotdist}
		\dot{d}=\int_{0}^{\widetilde{r}}\frac{\ud}{\ud t}\left[\sqrt{g_{i\:\!k}\frac{\ud x^i}{\ud l}\,\frac{\ud x^k}{\ud l}}\right]\,\ud l\,.
	\end{equation}
    Since the metric tensor $g_{i\:\!k}$ depends on time upon the known functions, we can calculate the integrals (\ref{eqn:dist},\ref{eqn:dotdist}) numerically.
	
	To evaluate the local value of $H_0$, we take the sample of 389 Ia supernovae from the \emph{Cosmic Flows 3} supernovae data \cite{2009AJ....138..323T}. For each object, we set $\widetilde{r}$ as its luminosity distance and $(\theta,\phi)$ as its angular position. The only parameter, which remains unspecified yet, is the parameter $\lambda$ proportional to the amplitude of the inhomogeneities. We relate $\lambda$ to the new parameter $\Omega_{Virgo}=\lambda\,\rho^{(1,0)}/\rho_{crit}$ defined as the density above the average value $\Omega_m$, evaluated at the center of the Virgo cluster and expressed in the critical density units. In our analysis, we consider different values of $\Omega_{Virgo}$. We calculate numerically the integrals (\ref{eqn:dist},\ref{eqn:dotdist}) to each supernova position and find the local Lema\^itre-Hubble constant by performing the linear fit to the $(d,\dot{d})$ results.
	
\begin{figure}[h]
		\centering
		\includegraphics[width=0.65\textwidth]{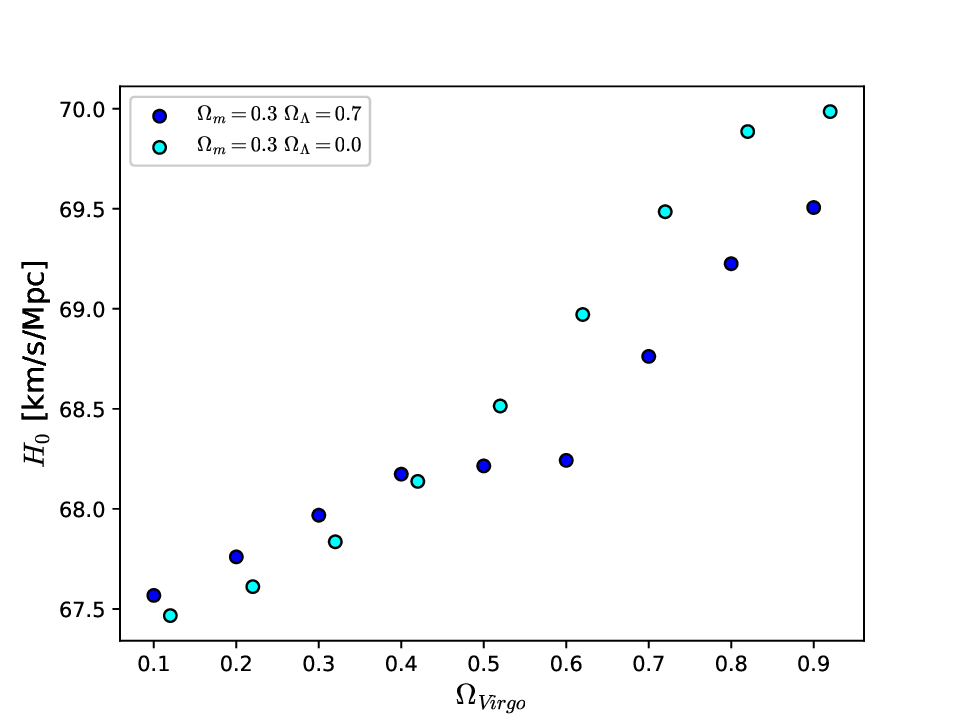}
		\caption{\label{fig:HubbleD} \scriptsize{The local value of the Lema\^itre-Hubble constant $H_0=\dot{d}/d$ for the two examplary background parameters $(\Omega_m,\Omega_\Lambda)$ and several values of the amplitude of the inhomogeneities labeled by the $\Omega_{Virgo}$ quantity described in the text}. }
	\end{figure}

	In the Figure \ref{fig:HubbleD}, we show the resulting local value of the Lema\^itre-Hubble constant $H_0$ for the two specific background models and various $\Omega_{Virgo}$ values. The Fourier representation of the $d_i(x,X_i,a_i)$ provides us a simple formula for the minimum density in the elementary cell:
	\begin{equation}
		\rho_{min}=\rho_0+3\lambda\,\frac{a(t)\,\dot{a}(t)\,\dot{\mathscr{A}}_{10}(t)-\alpha}{8\pi\,a^2(t)}\,\sum\limits_{i=1}^{S}A_i\,I_0(a_i)\,.
	\end{equation}
	Note again, that the fraction before the sum is negative, so $\rho_{min}$ is lower than the background density $\rho_0$. The condition that the density should be always positive, so that $\rho_{min}>0$, gives us an upper bound of the amplitude $\Omega_{Virgo}$. For the parameter $\Omega_m=0.3$, this upper bound is $\Omega_{Virgo}=0.97$. For that reason, in the Figure \ref{fig:HubbleD} we restrict the values of the amplitude of the inhomogeneities up to $\Omega_{Virgo}$ around 0.9. Another issue is to find the value of $\Omega_{Virgo}$, for which the amplitude of inhomogeneities becomes too high to be satisfactorily described by the second-order perturbation theory, and the higher orders are needed to preserve the energy-momentum tensor of the dust matter with a desired accuracy. We will address this question in subsection \ref{sec:2dlimits}.
	
	As we can see in the Figure \ref{fig:HubbleD}, for both background models considered, the value of the constant $H_0$ is higher than the CMB-based value $H_0=67.4\,\mathrm{km/s/Mpc}$ and there appears a positive correlation between the amplitude of the inhomogeneities and the value of $H_0$. Although such behavior suggests that inhomogeneities could influence the expansion of the universe, one cannot compare the Lemaitre-Hubble constant defined as $H_0=\dot{d}/d$ with the observations, because we never measure a distance on a hypersurface of a constant time, we measure it down our past light cone. The difference between $\dot{d}/d$ and the background value of $H_0$ reflects only the fact that the expansion scalar $\nabla_\mu\,U^\mu$ of the considered model is not homogeneous in space.
	In the real measurements, like e.g. \cite{2022ApJ...934L...7R}, the Lemaitre-Hubble constant is inferred from the relation between luminosity distance and redshift. We analyze this relation in the framework of our model in the next subsection.
	
	\subsection{Luminosity distance vs redshift \label{Sec:Luminosity}}
	In the standard FLRW geometry, the luminosity distance:
	\begin{equation}
		d_L = \left(\frac{\mathcal{L}}{4\pi \mathcal{F}}\right)^{1/2} \;,
	\end{equation}
	where $\mathcal{L}$ and $\mathcal{F}$ are the intrinsic luminosity and observed flux, can be expressed as a function of redshift and cosmological parameters through (see e. g. \cite{1998AJ....116.1009R}):
	\begin{equation}
		d_L = cH_0^{-1}(1+z)|k|^{-1/2}\mathcal{S}_k \left[|k|^{1/2}\int_0^z \mathrm{d}z\left(\left(1+z\right)^2\left(1+\Omega_m^2\right)-z\left(2+z\right)\Omega_{\Lambda}\right)^{-1/2}\right] \;,
	\end{equation}
	where $\mathcal{S}_k(\xi):=\mathrm{sinh}\,\xi$ for $k<0$, $\mathcal{S}_k(\xi):=\mathrm{sin}\,\xi$ for $k>0$ and in the case of the flat universe $\mathcal{S}_k(\xi)=\xi$.
	
	We are primarily interested in the supernovae observations performed at the relatively low redshifts and thus expansion around $z=0$ (necessarily for $z<1$) are usually considered reliable for such data sets. In this context, the cosmographic Hubble law can be used, with the Taylor-expanded formula reading \cite{2004Visser}:
	\begin{eqnarray}\label{eqn:dLTaylor}
		d_L &=& \frac{cz}{H_0}\Bigl( 
		1+\frac{1}{2}\left(1-q_0\right)z \nonumber \\&-&\frac{1}{6}\left(1-q_0-3q_0^2+j_0+\frac{kc^2}{H^2_0}a^2_0\right)z^2 +O(z^3)\Bigr)\;,
	\end{eqnarray}
	where:
	\begin{equation}
		H \equiv \frac{\dot{a}}{a}\;\;,\;\;q\equiv-\frac{\ddot{a}}{aH^2}\;\;,\;\;j \equiv \frac{\dddot{a}}{aH^3} \;,
	\end{equation}
	are Hubble parameter, deceleration and jerk respectively. For clarity, we keep explicitly the speed of light in these formulas, while in other places in the text we use the natural units $c=1$ as indicated previously.
	
	One of the hypotheses explaining the Hubble tension problem states that the local inhomogeneities could affect the observed $d_L(z)$ relation. When one fits the formula (\ref{eqn:dLTaylor}), valid in the homogeneous case, to the $d_L(z)$ data influenced by the matter inhomogeneities, the resulting $H_0$ value could differ from the FLRW prediction. To test this hypothesis, for given values of the background parameters $(\Omega_m,\Omega_\Lambda)$ and the fixed value of the amplitude $\Omega_{Virgo}$, we generate a mock catalog of sources and the $d_L(z)$ diagram and fit the $H_0$ as the free parameter in (\ref{eqn:dLTaylor}). The procedure for creating the mock catalog is the following.

	First, we formulate the geodesic equation and the focusing equation \cite{2004LRR.....7....9P} without shear as the first-order differential equations (see \cite{2005Barausse} for analytical expressions):
	\begin{equation}\label{eqn:Rk4}
	\frac{\ud x^\mu}{\ud l}=k^\mu\quad,\quad \frac{\ud k^\mu}{\ud l}=-\Gamma^{\mu}_{\alpha\beta}\,k^\alpha\,k^\beta\,,\quad
	\frac{\ud d_A}{\ud l}=\dot{d}_A\,,\quad
	\frac{\ud\dot{d}_A}{\ud l}=-\frac{1}{2}\,\left(R_{\alpha\:\!\beta}\,k^\alpha\:\!k^\beta\right)\,d_A\,		\qquad \qquad
	\end{equation}
 where $k^{\mu}$ is the wave vector, $d_A$ is the angular diameter distance and $R_{\alpha \beta}$ is the Ricci tensor. 
 Neglecting shear significantly simplifies equations and can be justified as it does not affect the redshift-distance relation apart from the strong lensing cases \cite{Bolejko2016}.
	In this subsection, we denote the affine parameter along the geodesic as $l$, to avoid confusion with the perturbative parameter $\lambda$. We integrate these equations by using the fourth-order Runge-Kutta method with the following initial conditions. We locate the observer initially at the Milky Way position at the universe's age $x^\mu(l=0)=(t_0,0,0,0)$. We fix the initial wave vector as the past oriented vector with the fixed $k^0=-1$. We generate randomly its spatial direction $\vec{k}$ with the probability distribution uniform on the unit sphere and then normalize the length of a three-vector $|\vec{k}|$ so that the initial wave vector is null $k^\mu\,k_\mu=0$. The preservation of the null condition $k^\mu\,k_\mu=0$ helps us to establish a proper Runge-Kutta step during the numerical integration. Initially, the angular diameter distance is $d_A(l=0)=0$, and its derivative equals $\dot{d}_A(l=0)=1$. The result of numerical integration of (\ref{eqn:Rk4}) is the past oriented null geodesic $x^\mu(l)$, the corresponding wave vector $k^\mu(l)$, the angular diameter distance along the geodesic $d_A(l)$ and its derivative $\dot{d}_A(l)$. Assuming that the light emitter and the observer are both comoving with matter, and have the four-velocity $U^\mu=(1,0,0,0)$, we can express the redshift along the geodesic as $z(l)=-k^0(l)-1$. Next, we derive the function $d_L(z)$ along the geodesic using the reciprocity relation $d_L=(1+z)^2\,d_A$ between the luminosity distance $d_L$ and the angular diameter distance $d_A$.
	
\begin{figure}[h]
		\centering
		\begin{tabular}{cc}
			\includegraphics[width=0.47\linewidth]{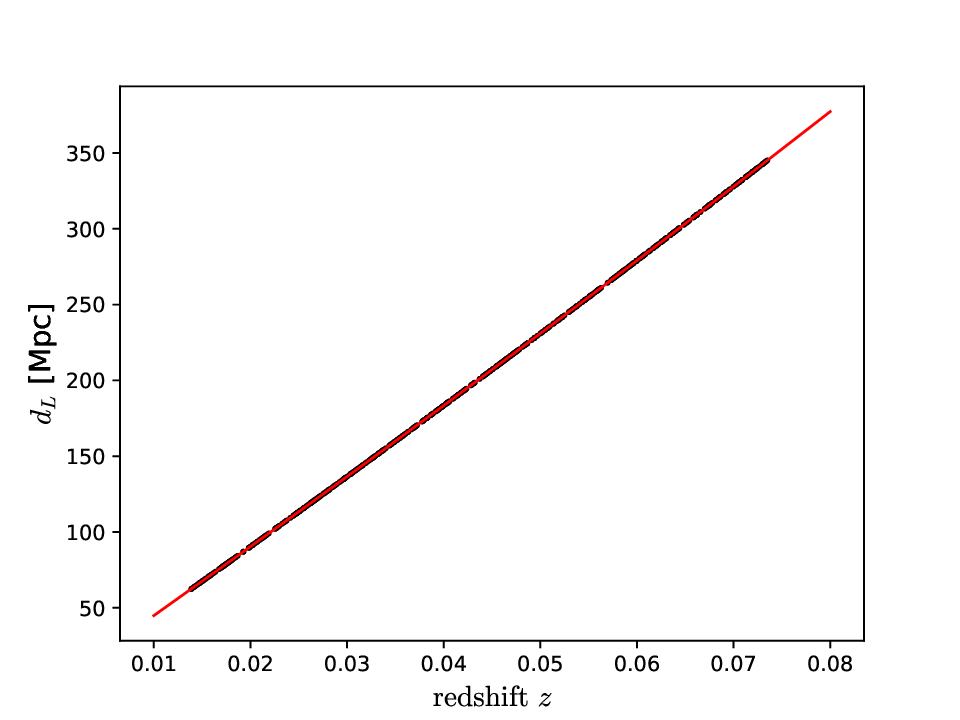}&
			\includegraphics[width=0.47\linewidth]{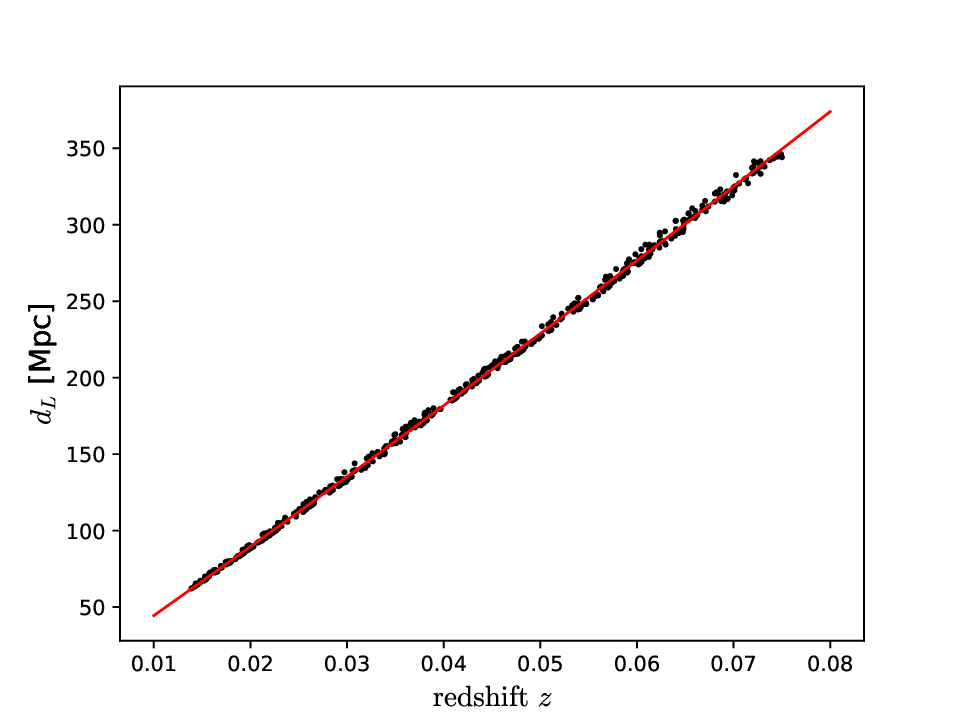}\\
			\includegraphics[width=0.47\linewidth]{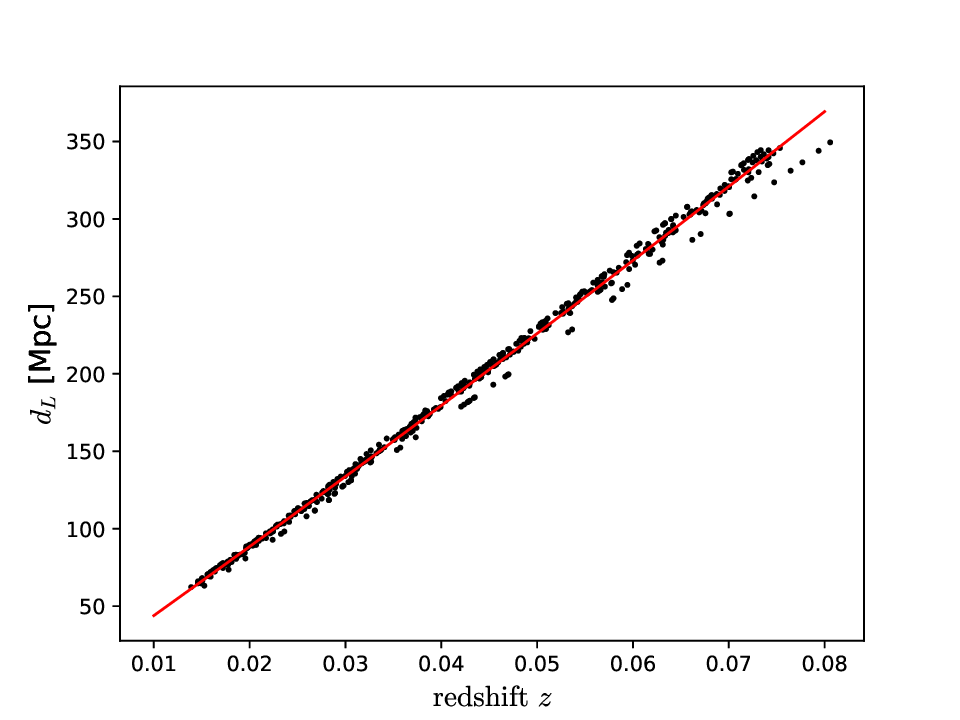}&
			\includegraphics[width=0.47\linewidth]{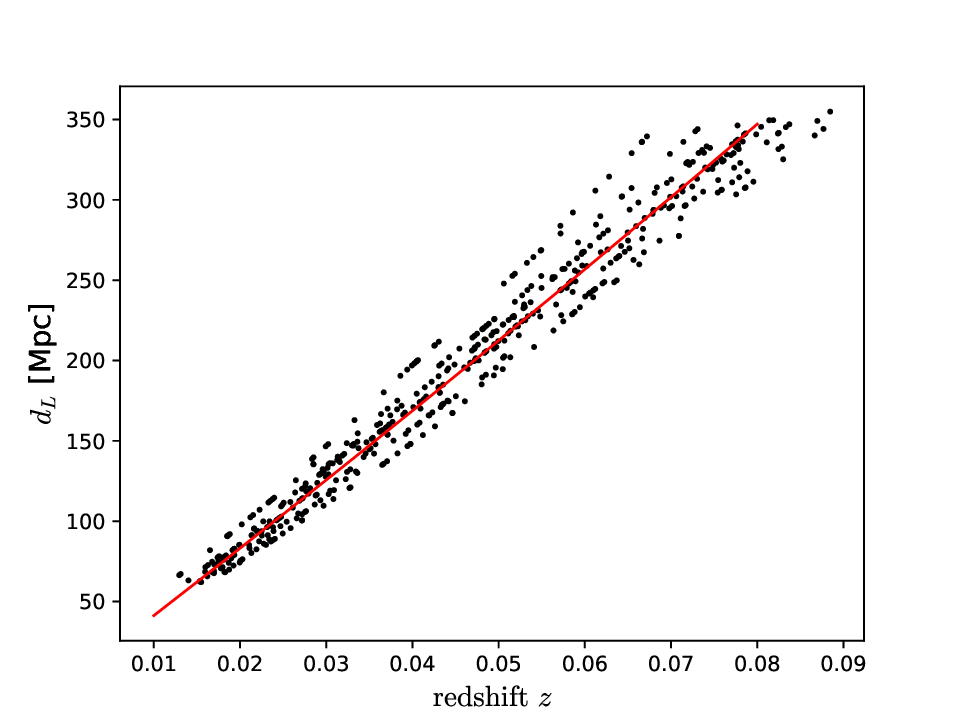}
		\end{tabular}
		\caption{\label{fig:diagram4}\scriptsize{The $d_L(z)-z$ relation for $\Omega_m=0.3$, $\Omega_\Lambda=0.7$ and various amplitudes: $\Omega_{Virgo}=0.0$ - \emph{top left}, $\Omega_{Virgo}=0.25$ - \emph{top right}, $\Omega_{Virgo}=0.3$ - \emph{bottom left}, and $\Omega_{Virgo}=0.4$ - \emph{bottom right}.}}
	\end{figure}

	In the inhomogeneous universe, the relation $d_L(z)$ is in principle anisotropic and depends on the direction of the initial wave vector $\vec{k}$. Therefore, after fixing the background parameters and the amplitude of the inhomogeneities, we generate ten geodesics with random initial directions. From each geodesic, we randomly select fifty points corresponding to the angular diameter distance higher than the size of the elementary cell. In effect, we obtain the $d_L(z)$ diagram with 500 points. We present in Figure \ref{fig:diagram4} four such diagrams generated for the $(\Omega_m=0.3,\Omega_\Lambda=0.7)$ background and different $\Omega_{Virgo}$ parameters.
	
\begin{figure}[h]
		\centering
		\includegraphics[width=0.65\textwidth]{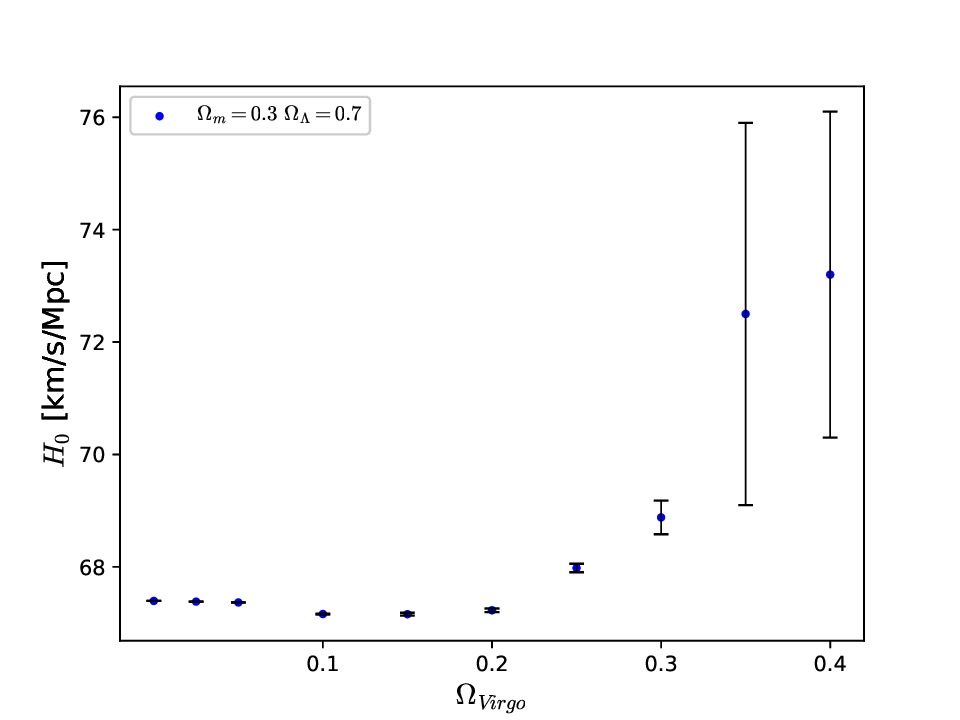}
		\caption{\label{fig:H037results} \scriptsize{The $H_0$ as a function of $\Omega_{Virgo}$ for the background model with the parameters $\Omega_m=0.3$ and  $\Omega_{\Lambda}=0.7$.}}
	\end{figure}

\begin{figure}[h]
		\centering
		\includegraphics[width=0.65\textwidth]{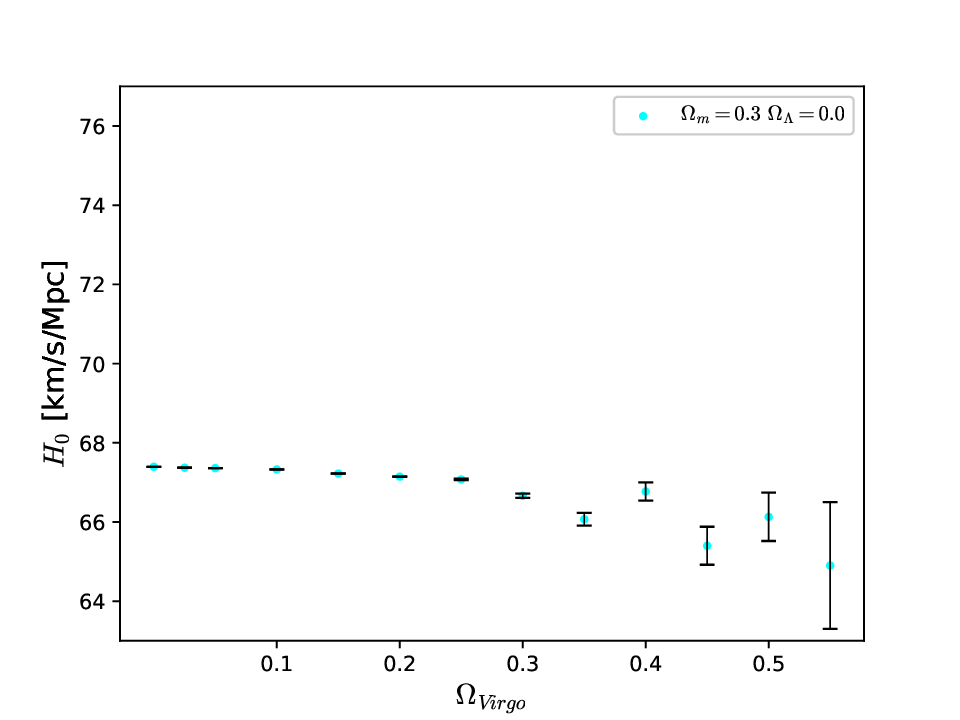}
		\caption{\label{fig:H030results} \scriptsize{The $H_0$ as a function of $\Omega_{Virgo}$ for the background model with the parameters $\Omega_m=0.3$ and  $\Omega_{\Lambda}=0.0$.}}
	\end{figure}

	In the Taylor-expanded formula (\ref{eqn:dLTaylor}), we fix the background values of the deceleration parameter $q_0$ and the jerk parameter $j_0$, and then we fit $H_0$ by the least-squares method applied to the $d_L(z)$ diagram. We plot the results for the $\Lambda$CDM background ($\Omega_m=0.3$, $\Omega_{\Lambda}=0.7$) as a function of the amplitude $\Omega_{Virgo}$ in Figure \ref{fig:H037results}. The results of the similar analysis without the cosmological constant ($\Omega_m=0.3$, $\Omega_{\Lambda}=0.0$) are plotted in Figure \ref{fig:H030results}.
		One can observe that the points on the $d_L(z)$ diagram are scattered with a variance positively correlated with the amplitude of the inhomogeneities. The uncertainty of the local measurements of the Lemaitre-Hubble constant is of order of $1\,\mathrm{km/s/Mpc}$ \cite{2022ApJ...934L...7R}. When the amplitude parameter $\Omega_{Virgo}$ is sufficiently large, the standard deviation of $H_0$, estimated by the least-squares method on the mock data, becomes comparable to or greater than the uncertainty in the real measurements. Moreover, for the amplitude $\Omega_{Virgo}$ large enough, the estimated $H_0$ presented on both figures significantly differs from the background value $H_0=67.4\,\mathrm{km/s/Mpc}$. For the $\Lambda$CDM background, the estimated values of $H_0$ are consistent with the value inferred from the distance ladder measurements $H_0=73.04\,\mathrm{km/s/Mpc}$ \cite{2022ApJ...934L...7R}. For the background without the cosmological constant, the effect is opposite, and the estimated $H_0$ is slightly lower than the CMB-based value. We conclude that, in the presence of the inhomogeneities, there exist a significant discrepancy between the background value of the Lema\^itre-Hubble constant and the $H_0$ estimated as the best fit to the $d_L(z)$ relation (\ref{eqn:dLTaylor}). This effect depends on the amplitude of the inhomogeneities and on the background dynamics. 

	\subsection{Limitations of the second-order perturbation theory}\label{sec:2dlimits}
	A characteristic feature of the perturbation theory is that the perturbed metric $g_{\mu\:\!\nu}=g^{(0)}_{\mu\:\!\nu}+h_{\mu\:\!\nu}$ should be close to the background in the sense that $|h_{\mu\:\!\nu}|<1$. The second-order perturbations, considered here, enable us to describe a higher amplitude of perturbations than in the linear regime. However, a second-order solution also has its limitations.
	\begin{figure}[h]
		\centering
		\includegraphics[width=0.65\textwidth]{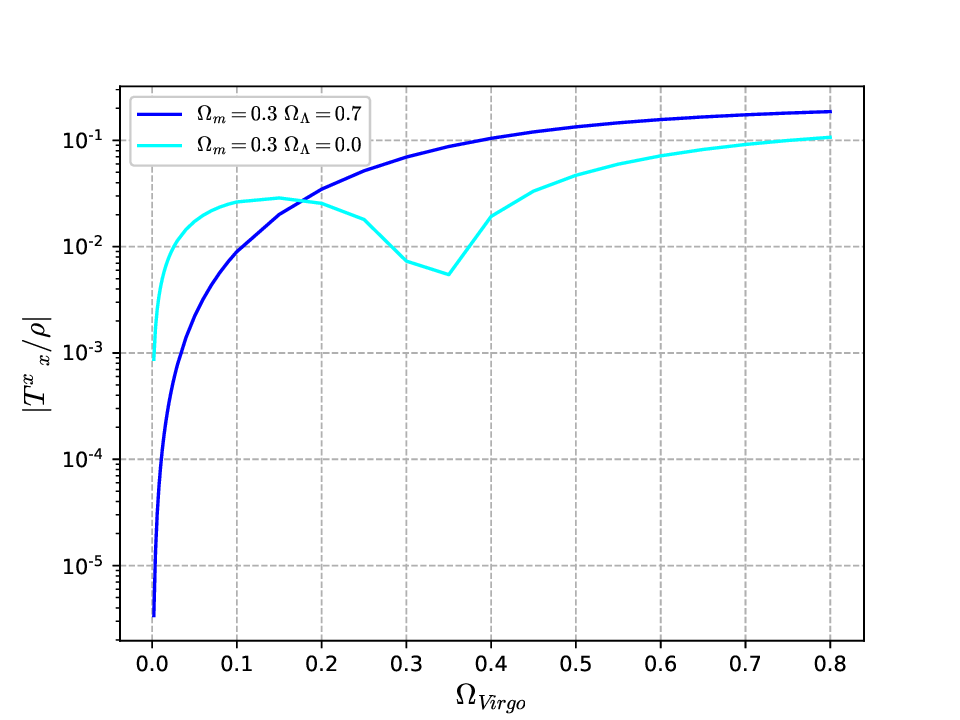}
		\caption{\label{fig:pressure} \scriptsize{The pressure component relative to the energy density at the center of the Virgo Cluster as a function of the inhomogeneities amplitude $\Omega_{Virgo}$.}}
	\end{figure} 
    In our previous work \cite{2023CQGra..40b5002S}, we have shown that the third and higher-order contributions to the energy-momentum tensor have non-diagonal elements, which are a few orders of magnitude smaller than diagonal elements. To check the reliability of the considered solution, we show in Figure \ref{fig:pressure} the diagonal pressure-like term $T^x\,{}_x$ relative to the energy density as a function of the amplitude $\Omega_{Virgo}$, evaluated at the center of the Virgo cluster. Figures presenting other pressure components $T^y\,{}_y$ and $T^z\,{}_z$ look the same. Unfortunately, the values of the inhomogeneities amplitude $\Omega_{Virgo}>0.3$, for which the results of the local measurements of the Hubble constant are the most interesting, correspond to the pressure-like contribution to the energy-momentum tensor of the order of a few percent of the energy density. Such values are too high to describe the energy-momentum tensor of the dust with reliable accuracy. Therefore, the results related to the Hubble tension, presented in previous subsections, should be understood as a kind of extrapolation, and further studies referring to the more physical energy-momentum tensor are still needed.
    
    \section{Conclusions}
    \label{Sec:Conclusions}
    In this work, we combine the cosmography of the local Universe with the perturbative solution of the Einstein equations up to the second order. The symmetry of the metric tensor enables us to simplify Einstein equations, without restricting significantly allowed density distributions. We constrain the metric functions related to the density distribution by using the individual galaxies data from the Extragalactic Distance Database. In effect, the metric tensor considered is consistent with the distribution of nearest galaxy clusters. Our model can be used in examining observable phenomena requiring realistic templates for local geometry and energy density e.g. the bulk flow (see e.g. \cite{Watkikns}). We show that inhomogeneities in the distribution of matter can significantly affect the $d_L(z)-z$ relation. When fitting the numerically obtained $d_L(z)$ for the perturbed metric to the formula (\ref{eqn:dLTaylor}), valid in the homogeneous FLRW case, the additional gravitational degrees of freedom become dressed. In particular, the fitted Lema\^itre-Hubble constant differs from its background value. This effect depends on the amplitude of inhomogeneities and the background parameters. In the case of the $\Lambda$CDM background, the estimated $H_0$ agrees with the Lema\^itre-Hubble constant value measured by the distance ladder methods. This result is, however, at the limit of applicability of the second-order perturbation theory, so further studies considering more reliable energy-momentum tensor are still required to better understand the persistent problem of the Hubble tension.
    \section*{ACKNOWLEDGMENTS}
    J. J. O. acknowledges the support of the National Science Centre
(NCN, Poland) under the Sonata-15 research Grant No. 2019/35/D/ST9/00342. 

Some analytical and numerical calculations are performed with the help of the Maxima Computer Algebra System and the Fortran 95 code.
%    \newpage
%\normalsize
\appendix
	
    \section{Second-order metric functions}\label{sec:2d_order_metric}
	Below, we present the explicit form of the metric tensor in the second order. For further details, we refer the Reader to our previous article \cite{2023CQGra..40b5002S}. 
	
	In the metric (\ref{eqn:Metric1}), we label the order of expansion by $l$. The second order with $l=2$ has three terms:
	\begin{eqnarray}\label{eqn:metric_20}
		c_{i\:\!j}^{(2,0)}=\left(\mathscr{A}_{20}(t)C_{20}(x^i)+\mathscr{D}_{20}(t)\sum\limits_{s=1}^{3}D_{20}(x^s)\right)\delta_{i\:\!j}+ \\
		\nonumber+\mathscr{F}_{20}(t)\,F_{20}(x^i,x^j)\,(1-\delta_{i\:\!j})\,,
	\end{eqnarray}
    \begin{eqnarray}\label{eqn:metric11}
    	c_{i\:\!j}^{(1,1)}=\left(\mathscr{A}_{11}(t)C_{11}(x^i)+\mathscr{D}_{11}(t)\sum\limits_{s=1}^{3}D_{11}(x^s)\right)\delta_{i\:\!j}+ \\
    	\nonumber+\mathscr{F}_{11}(t)\,F_{11}(x^i,x^j)\,(1-\delta_{i\:\!j})\,,
    \end{eqnarray}
    and
    \begin{eqnarray}
    c_{i\:\!j}^{(0,2)}=\frac{3}{16}\widetilde{r}\,{}^4\,\delta_{i\:\!j}\,.
    \end{eqnarray}
	The Taylor-expanded up to the second order Einstein equations are approximately satisfied by the dust-like energy-momentum tensor, if only the metric functions introduced above fulfill the following conditions:
	\begin{equation}\label{eqn:D20}
		\mathscr{D}_{20}(t)=a^2(t)\,\dot{\mathscr{A}}_{10}(t)^2\,, 
	\end{equation}
	\begin{equation}\label{eqn:F20}
		\mathscr{F}_{20}(t)=\frac{1}{4}a^2(t)\,\dot{\mathscr{A}}_{10}(t)^2+\alpha\,\mathscr{A}_{10}(t)\,,
	\end{equation}
    \begin{equation}\label{eqn:A2}
    	a^2(t)\,\ddot{\mathscr{A}}_{20}(t)+3\,a(t)\,\dot{a}(t)\,\dot{\mathscr{A}}_{20}(t)=2\,\alpha\,\mathscr{A}_{10}(t)\,,
    \end{equation}
	\begin{equation}\label{eqn:spatialD20}
	%\frac{\ud^2}{\ud w^2}D_{20}(w)=-\frac{1}{2}\left(\frac{\ud^2}{\ud w^2}C_{10}(w) \right)^2\,,
	D''_{20}(w)=-\frac{1}{2}\,C''_{10}(w)^2\,,
    \end{equation}
    \begin{equation}\label{eqn:spatialF20}
	%\frac{\partial^2}{\partial w\partial v}F_{20}(v,w)=-\frac{\ud^2}{\ud v^2}C_{10}(v)\,\frac{\ud^2}{\ud w^2}C_{10}(w)\,,
	\frac{\partial^2}{\partial w\partial v}F_{20}(v,w)=-C''_{10}(v)\,C''_{10}(w)\,,
    \end{equation}
    \begin{equation}\label{eqn:spatialC20}
	C_{20}(w)=-\frac{1}{4}\,C'_{10}(w)\,C'''_{10}(w)\,,
    \end{equation} 
    \begin{equation}\label{eqn:D11}
    	\mathscr{D}_{11}(t)=\mathscr{F}_{11}(t)=\mathscr{A}_{10}(t)\,,
    \end{equation} 
    \begin{equation}\label{eqn:A11}
    	a^2(t)\,\ddot{\mathscr{A}}_{11}(t)+3\,a(t)\,\dot{a}(t)\,\dot{\mathscr{A}}_{11}(t)=\mathscr{A}_{10}(t)\,,
    \end{equation}
    \begin{equation}\label{eqn:spatialD11}
    	D_{11}(w)=C_{10}(w)-\frac{1}{2}\,w\,C'_{10}(w)\,,
    \end{equation}
    \begin{equation}\label{eqn:spatialF11}
    	F_{11}(v,w)=\frac{1}{2}\left(v\,C'_{10}(w)+w\,C'_{10}(v) \right)\,,
    \end{equation}
    \begin{equation}\label{eqn:spatialC11}
    	C_{11}(w)=C''_{10}(w)\,,
    \end{equation}
    where $v,w\in \{x,y,z\}$. When the scale factor $a(t)$ and the first-order function $\mathscr{A}_{10}(t)$ are known, the functions $\mathscr{D}_{20}(t)$, $\mathscr{F}_{20}(t)$, $\mathscr{D}_{11}(t)$ and $\mathscr{F}_{11}(t)$ are explicitly given by (\ref{eqn:D20}, \ref{eqn:F20}, \ref{eqn:D11}). The functions $\mathscr{A}_{20}(t)$ and $\mathscr{A}_{11}(t)$ are solutions of differential equations (\ref{eqn:A2}, \ref{eqn:A11}). These differential equations are of the same type as the equation (\ref{eqn:A1}). The remaining functions of spatial variables: $D_{20}(x)$, $D_{20}(y)$, $D_{20}(z)$, $F_{20}(x,y)$, $F_{20}(y,z)$, $F_{20}(x,z)$, $C_{20}(x)$, $C_{20}(y)$, $C_{20}(z)$, $D_{11}(x)$, $D_{11}(y)$, $D_{11}(z)$, $F_{11}(x,y)$, $F_{11}(y,z)$, $F_{11}(x,z)$, $C_{11}(x)$, $C_{11}(y)$, $C_{11}(z)$ can be found from (\ref{eqn:spatialD20}, \ref{eqn:spatialF20}, \ref{eqn:spatialC20}, \ref{eqn:spatialD11}, \ref{eqn:spatialF11}, \ref{eqn:spatialC11}), if only the functions $C_{10}(x)$, $C_{10}(y)$, $C_{10}(z)$ are given. Therefore, specification of the background and the functions $C_{10}$, which we relate to the observed distribution of galaxies in Sec. \ref{Sec:dens_distr}, completely determine the metric tensor in the second order.

	\bibliography{References2023}
	\bibliographystyle{iopart-num}

\end{document}